\documentclass{appolb}
\usepackage{epsfig}
\usepackage{amsmath}
\begin{document}
\title{On the Geometry of Random Networks%
\thanks{Talk at the Workshop on Random Geometries, 
Krakow, May 15-17, 2003. To appear in the Proceedings, to be
published in Acta Physica Polonica B.}%
}
\author{A. Krzywicki
\address{Laboratoire de Physique Th\'eorique, B\^atiment 210, 
Universit\'e Paris-Sud, 91405 Orsay, France}
}
\maketitle
\begin{abstract}
The Krakow-Orsay collaboration has 
applied methods borrowed from
equilibrium statistical mechanics 
and analytic combinatorics to
study the geometry of random 
networks. Results contained in a series 
of recent publications and 
concerning networks that are either 
uncorrelated or causal are briefly overviewed.
\end{abstract}
\par LPT Orsay 03/46 
  
\section{Introduction}
The purpose of this communication 
is to overview the results published
recently in the Physical Review E 
by the Krakow-Orsay collaboration
\cite{krz1,krz2,krz3} and devoted 
to the geometry of random 
networks\footnote{Excellent general 
reviews on network physics can 
be found in ref. \cite{badm}.}. 
The discussion is sketchy 
and aimed to give the reader 
only a general idea of what 
has been achieved. All the 
useful details can be found 
in the original papers. 
\par
Network study is not our original 
field of research: we are trying
to exploit the experience gained 
working on a different subject -  
quantum gravity, or, more precisely, 
statistical mechanics of random 
manifolds - in another context and 
to fill the gap between two 
communities, that are differently 
motivated but often confronted to 
manifestly similar problems. This 
is a status report, in the same 
vein as my talk at the Utrecht 
symposium in 2001, addressed to a 
similar audience \cite{krz4}. 
\par
There are two complementary 
approaches to random networks, and actually
to numerous complex systems: the 
diachronic and the synchronic one.
In the former one focuses on the 
time evolution of the system. It is
particularly suitable if the aim is 
to uncover the evolution dynamics.
In the latter one works at a fixed 
time, considering an ensemble
of related systems, with the aim 
of finding common structural traits.
Being primarily interested in the 
geometry of networks, we have adopted 
the synchronic approach.
\par
The goal is to develop a statistical 
mechanics of random networks. 
In the statistical mechanics 
of gases one starts with an ideal gas. 
Analogously, working with uncorrelated 
nodes is a natural first step 
in our research. For definiteness, we 
consider undirected networks only.

\section{Networks with uncorrelated nodes}
\subsection{Formulation of the model}
The model is compactly defined 
by writing the partition function as a 
formal integral
\begin{equation}
Z \sim \int_{-\infty}^{+\infty} d\phi \;
\exp{\frac{1}{\kappa}[-\phi^2/2\lambda + 
\sum_{n>0} p_n \phi^n]}
\label{1}
\end{equation} 
The set of "coupling constants" $p_n$ is 
eventually identified with the 
degree distribution, while $\kappa$ and 
$\lambda$ are control parameters. Of 
course, strictly speaking, the integral 
does not exist. However, the perturbative
series in the "coupling constants" is 
well defined and the individual terms
can be, as in field theory, represented 
by Feynman diagrams. The idea is
to identify the labeled Feynman diagrams of 
the minifield theory defined by 
(\ref{1}) with the graphs representing the 
network and to attach to every such 
graph a weight equal to the corresponding 
Feynman amplitude. All this is explained 
in detail in refs. \cite{krz1,krz2}. 
\par
Consider the ensemble where $N$ and $L$, the 
number of nodes and links, respectively, 
are fixed. Using the Feynman rules one finds 
that the weight $w$ of a labeled 
non-degenerate graph - \ie one without 
tadpoles and multiple connections between 
nodes - is up to an irrelevant factor given by
\begin{equation}
w \sim \prod_{j=1}^N n_j!\; p_{n_j}
\label{2}
\end{equation}
where $n_j$ is the degree of 
the $j$-th node. This graph is, in
general, not connected.
\par
It follows from the obvious identity
\begin{equation}
\sum_{j=1}^N n_j = 2L
\label{2bis}
\end{equation}
that, at fixed $N$ and $L$, all non-degenerate 
graphs are equiprobable when
$p_n$ has the Poisson form $p_n \sim c^n/n!$, 
with some constant $c$. Hence, 
in this case graphs are those of the classical 
Erd\"os-Renyi theory \cite{bol}. 
In general, the statistical ensemble discussed 
in this section is a generalization 
of the classical ensemble of random 
graphs to allow for an arbitrary degree
distribution (see later).
\par
Notice, that, because of (\ref{2bis}), the 
relative weights of microstates are
invariant under the transformation $p_n \to c^n\; p_n$. 
We shall see later that
the ambiguity is lifted when one fixes the ratio $L/N$.

\subsection{A few words on trees}
In the "quasi-classical" limit $\kappa \to 0$ only 
connected tree graphs contribute
to $W = \kappa \ln{Z}$. The integral in 
(\ref{1}) can be calculated using the 
saddle-point approximation. The saddle-point condition reads
\begin{equation}
\Phi = \lambda \sum_{n>0} n p_n\; \Phi^{n-1} 
\label{3}
\end{equation}
and one easily checks that $\Phi = \partial
 W/\partial p_1$, which means that 
$\Phi$ generates tree graphs with one 
external node marked. Although eq. (\ref{3}) 
can be exactly solved by Lagrange inversion, 
it is sufficient to use a more
direct approach \cite{adj}: Eq. (\ref{3}) can 
only be satisfied when $\lambda$ is
smaller than some critical value $\lambda_c$. 
Hence $\Phi$ is a singular function
of $\delta \lambda = \lambda_c - \lambda$. 
Furthermore, only the singular part 
of $\Phi$ is of real physical interest, 
since it determines the behavior of 
arbitrarily large trees. And this singular part 
is readily found directly from 
(\ref{3}). The result is used to determine 
the distribution of the smallest distance 
between pairs of nodes, the Hausdorff and 
spectral dimensions, etc. All this is
discussed at length in ref. \cite{krz1}, where, 
among others, the results of
ref. \cite{adj} are extended to the 
interesting case of scale-free graphs. I
shall not enter into more details here.

\subsection{Algorithmic considerations}
Equation (\ref{2}) gives a weight to 
each microstate. For given $L/N$, this, 
in essence, defines the statistical 
ensemble. However, in order to make this 
definition being of any use, we supplement 
it with an explicit recipe enabling 
one to construct graphs, \eg on a computer. 
To this end we define a local 
{\em move} transforming one graph into 
another. A succession of such moves 
is a Markov process. The initial state 
of the system is rapidly forgotten 
and graphs are sampled with relative 
frequency given by (\ref{2}). The 
whole procedure is a specific application 
of the so-called Metropolis 
method, widely used in other branches 
of statistical physics \cite{met}.
\par
One move consists of three steps \footnote{An 
equivalent and simpler definition is
given in refs. \cite{krz1,krz2}. The one given 
here makes the useful mapping on the 
balls-in-boxes model more evident (\cf footnote [15] 
in ref. \cite{krz2}.}. First,
we sample two distinct nodes, say $j$ and $k$. 
Second, we pick one neighbor of $j$,
say $i$. Third, we rewire $ij \to ik$ with probability
\begin{equation}
\mbox{\rm Prob}( ij \to ik) = \mbox{\rm min}\bigl(1, 
\frac{p_{n_k+1}\; p_{n_j-1}} {p_{n_k}\; p_{n_j}}\bigr)
\label{4}
\end{equation}
It is evident, that as far as the modifications 
of node degrees are concerned,
the algorithm is identical to that defining the 
so called balls-in-boxes model
\cite {bbj}, defined by the partition function
\begin{equation}
z \sim \sum_{\{n_i\}} \prod_{j=1}^N p_{n_j} 
\delta(M - \sum_{j=1}^N n_j)
\label{5}
\end{equation}
and descibing $M$ balls distributed with 
probability $\sim p_n$ among $N$ boxes (in
out case $M=2L$). The constraint represented 
by the Kronecker delta is in the
limit $N \to \infty$ satisfied "for free" by virtue 
of the law of large numbers
provided $M/N = \langle n \rangle \equiv \sum_n n
 p_n/\sum_n p_n$ . When the
last condition is met, the occupation number 
of one box $\to p_n$ when  
$N \to \infty$.
\par
The isomorphy of the graph and balls-in-boxes 
model implies that the degree
distribution in the graph model tends 
asymptotically to $p_n$ provided
\begin{equation}
L = \frac{1}{2} N \langle n \rangle
\label{5bis}
\end{equation}
Although the relative weigths of microstates are 
invariant under the 
transformation $p_n \to c^n\; p_n$, $\langle n \rangle$ 
is not. Hence, 
the ratio $L/N$ is fixed, once one 
has decided that the degree distribution 
should be $p_n$.
\par
There is a problem, however. Graphs generated 
by the above described algorithm 
are, in general, degenerate (as are the 
objects constructed in the well-known 
paper by Molloy and Reed \cite{mr}; these 
construction is often misused in the 
physics literature as a method of generating 
graphs, without due attention to 
the degeneracy problem).
\par
Our algorithm is local. The creation 
of degeneracies is therefore easily 
forbidden: It suffices to check that $i$ 
and $k$ are neither identical 
nor connected. But this check introduces a bias. 
The point is discussed in the
next subsection.

\subsection{Finite-size effects and the degeneracy problem}
The use of the Metropolis method guarantees 
that the degree distribution
approaches $p_n$ for large $N$, provided 
one has enough statistics, \ie
when $Np_n \gg 1$, even if one forbids degeneracies. However,
this condition is not satisfied when $p_n$ 
has a fat tail. Then, there are large
fluctuations in the tail and introducing 
a constraint can bias the sample. 
\par
Assume that $p_n \sim n^{-\beta}$ at large $n$. 
At finite $N$ the tail of $p_n$
cannot extend to infinity, because there exists 
some $n_{max}$ such that the
expected number of nodes with degrees $n > n_{max}$ is 
less than unity. Neglecting
all correlations one easily finds the scaling law
\begin{equation}
n_{max} \sim N^{1/(\beta-1)}
\label{6}
\end{equation}
It is easily seen that $np_n$ can actually be very 
small well below this natural
cut-off.
\par
The bias associated with rejecting 
degeneracies can be evaluated \cite{krz2}.
Consider the symmetric adjacency matrix $C_{ij}$: 
the elements of say the $m$-th
row sum up to $n$, the degree of the $m$-th node. 
These elements equal either 0 
or 1 when the graph is non-degenerate, they 
are just positive integers when the graph 
is degenerate. We wish to compare the 
number of ways the $m$-th node can be
connected to $n$ other nodes, when one 
accepts or rejects degeneracies. The
problem reduces to counting the number of 
ways to place $n$ balls in $N-1$
boxes, but is not altogether trivial, since 
one has to take into account the
symmetry factors that appear in the weights 
of the degenerate graphs as well
as the shape of the degree distribution. 
The result at large $N$ is
\begin{equation}
\frac{\# \mbox{\rm without degeneracy}}
{\# \mbox{\rm with possible degeneracies}} 
\sim \exp[-\mbox{\rm const}\, \frac{n^2}{N}]
\label{7}
\end{equation}
Notice, that although $n/N$ is always small, $n^2/N$ 
may be large. We observe, that
at fixed $n$ the rejection of degeneraties does 
not introduce any bias at asymptotic
$N$. However, at large $n$ the rejection of 
degeneracies introduces a non-uniform
deformation of the spectrum. Actually, there is 
a cut-off scaling like $\sqrt{N}$.
This cut-off is smaller that the "natural" 
cut-off given by (\ref{6}) when
$2 < \beta < 3$. And this is not a marginal 
case. The $\beta$ exponent is like
that for most interesting networks! Apparently, 
forbidding degeneracies introduces 
a kind of "kinematic" correlation 
at finite $N$. It is important to
stress that this is a property of the model, 
not a deficiency of the algorithm.
Let us also mention, that the conclusions 
of the above heuristic argument are
confirmed by numerical simulations.
\par
There is a mathematical conclusion of the above 
discussion: the alorithm is fine.
To my knowledge this is the only efficient 
algorithm generating non-degenerate graphs
with a given degree distribution\footnote{We are, 
of course, ready to share our 
numerical code with interested people.}.
\par
There are also physical conclusions: Independently of 
any specific model, inter-node
correlations are necessarily present 
in observed scale-free networks, where the
tail of the degree distribution 
manifestly extends beyond a cut-off scaling
like $\sqrt{N}$. Also, the thermodynamical 
limit can be rather tricky for 
scale-free networks.

\subsection{Recent results by other people}
I would like to mention a very nice 
result obtained by Fronczak {\em et al} 
\cite{ffh}. They have calculated 
analytically the average internode distance 
in graphs with uncorrelated nodes:
\begin{equation}
\langle \mbox{\rm shortest path} \rangle \sim
\frac{\ln{N}}{\ln{(\langle n^2\rangle/\langle n\rangle -1)}}
\label{8}
\end{equation}
This formula has been proposed earlier, by other groups, but 
the derivation has
never been satisfactory, in my opinion. 
The problem is that the average shortest 
path has to grow like a power of $N$ 
for a generic tree with uncorrelated nodes 
\cite{adj}. Thus, a derivation leading 
to the logarithmic behavior must use 
arguments that do not work for trees. 
This condition is satisfied in ref. 
\cite{ffh}, but not in earlier 
publications claiming the same result. 
Notice that the coefficient in front of $\ln{N}$ diverges at the
percolation threshold, \ie when $\langle n(n-2)\rangle \to 0^+$
(\cf the celebrated reference \cite{mr}), at the 
transition to the regime dominated by trees.
\par
Another set of related and interesting 
results is presented in ref. \cite{bb}.
These authors have calculated, among others, 
the distribution of connected 
components and found the size of 
the percolation cluster above the 
percolation threshold. They have 
also calculated the conditional degree 
distribution of nodes belonging to the percolation cluster.
\par
There are many other results of the 
classical theory that could be
extended to graphs with a given 
degree distribution. Indeed, a comprehensive
discussion of the classical theory 
is a subject of a fat book \cite{bol}.
But, we feel we have understood some 
of the most salient features of the 
model without correlations. Also, 
we have a numerical control of the model.
Hence, we are eager to move to the 
next item on our agenda, \ie the
problem of correlations.

\section{Correlations}
A comprehensive theory of correlation 
in networks does not exist. It is 
straightforward to generalize the model of 
the preceding section, introducing 
pairwise correlations between degrees 
of neighbor nodes. Specific proposals 
to this effect have been made, for example, 
in refs. \cite{dor,bl}. However, 
it seems to me, that correlations of 
a different nature are particularly
important from the phenomenology point of view:
\par
- Correlations induced by the growth dynamics.
\par
- Clustering, \ie the fact that neighbors 
of a randomly chosen node are 
directly linked to each other more frequently than by chance. 
\par
A work on clustering is in progress, but 
we do not have yet results significant
enough to be presented here. Let me only 
mention that we are dealing 
with a very
specific class of matrix models. On the other hand, we 
have developed a synchronic
approach to growth processes, which is I believe worth mentioning:
\par
We focus our attention on trees, actually 
on labeled rooted trees, in order to 
be able to proceed analytically. We consider 
a static ensemble, but assume that 
the networks are endowed with a causal 
structure. We say a tree is endowed
with a causal structure when the labels 
always appear in growing numerical
order as one moves along the tree from 
the root towards an arbitrary node.
One can imagine that these labels refer 
to the time of node formation. The
approach is complementary to the more 
standard diachronic one. It turns out
that the presence of a causal structure 
generates internode correlations, once 
one has summed over all possible labelings. 
It is, therefore, of interest to 
consider models where these specific 
correlations do not interfere with 
correlations of a different origin. Hence, 
we assume that microstate 
weights factorize, as in eq. (\ref{2}). 
I have no place to enter into 
details, which can be found in ref. 
\cite{krz3} (see also the talk by 
P. Bialas \cite{bia}). Let me shortly 
summarize the most significant
results:
\par
- Some of the most popular growing 
network models, like Barabasi-Albert's 
\cite{ba}, can be reformulated in 
our static formalism. The original results
are recovered in an elegant fashion. 
This shows that the widely accepted
distinction between growing and 
equilibrium networks is not really correct.
The opposition between diachrony and 
synchrony is to large extent an illusion,
except when one is interested is specific 
phenomena, like aging, intrinsically 
reflecting the running of time.
\par
- We derive a closed, general formula for the degree distribution.
\par
- We also derive a closed formula for the 
correlation between the degree of 
an ancestor and that of its descendent, when 
they are separated by a geodesic 
distance $r$. Typically, the average descendent 
degree falls like $1/r$ \cite{jk}. 
Manifestly, this implies a long-range correlation.
\par
- We further derive a general formula 
for the distribution of the shortest
paths connecting nodes to the root. Using this formula 
we show that, generically,
the lenght of an average such path grows at most 
like $\ln{N}$, in contrast to the
uncorrelated trees where the 
growth is power-like \cite{adj,krz1}.

\section{Concluding remarks}
I am tempted to share with you a speculation, 
which does not rest on any 
solid basis, but may animate someone's 
imagination. Most present works on 
networks can be classified under the following headlines:
\par
- Geometry of networks.
\par
- Phenomenology of networks observed in nature.
\par
- Matter on quenched random networks (this 
includes \eg Ising spins living on 
networks, or the propagation of diseases).
\par
What is manifestly missing, as far as I know, 
is a study of networks whose geometry 
is interacting with matter living on it (like 
in the models of quantum gravity, we 
have been working on). I am not sure that it would be 
relevant for the present day
phenomenology, although some experts tell me that 
it might find applications in
the theory of traffic and communication. 
Nevertheless, I believe it would be
interesting to develop, at least, some models of that kind. 
I am pretty sure they
would find applications in the future.

\par
This work was partially supported by 
the EC IHP Grant No. HPRN-CT-1999-000161.
Laboratoire de Physique Th\'eorique 
is Unit\'e Mixte du CNRS UMR 8627.

\end{document}